\documentclass{cjour}
 \usepackage{psfig}

 \def\beq{\begin{eqnarray}}
 \def\eeq{\end{eqnarray}}
 
 \def\nns{\nonumber\\*}

\begin{document}

\title{A heap-based algorithm for the study of one-dimensional particle systems}
\author{Alain~Noullez$^{\dagger}$, Duccio~Fanelli$^{\ddagger}$,
Erik~Aurell$^{\ddagger,\star}$}

\titlerunninghead{Heap-based algorithm for one-dimensional particle systems}
\authorrunninghead{Noullez, Fanelli and Aurell}

\affil{
$^{\dagger}$  CNRS, Observatoire de la C\^ote d'Azur,\\
B.P. 4229, F-06304 Nice Cedex 4, France\\
$^{\ddagger}$  Dept. of Numerical Analysis and Computer Science,\\
KTH, SE-100 44 Stockholm, Sweden\\
$^{\star}$  Dept. of Mathematics, Stockholm University,\\
SE-106 91 Stockholm, Sweden
}

\abstract{
\noindent A fast algorithm to study one-dimensional self-gravitating systems,
and, more generally, systems that are Lagrangian integrable between collisions,
is presented.  The algorithm is event-driven, and uses a heap-ordered set of
predicted future events.  In the limit of large number of particles~$N$, the
operation count is dominated by the cost of reordering the heap after each
event, which goes asymptotically as~$\log N$.  Some applications are discussed
in detail.
}

\begin{article}

\section{Introduction}
\label{s:introduction}

\noindent In this paper we discuss a fast algorithm which integrates
numerically a one-dimensional system of $N$~interacting particles, provided the
dynamics can be Lagrangian integrated between two successive collisions.  An
important application is self-gravitating systems, since the gravitational
force is Lagrangian invariant in one dimension.  Several similar models with
Lagrangian invariant or quasi-invariant force fields can also be treated.

By computing all possible collision times between particles, we can select the
smallest of these and let the particles evolve until this time is reached.  The
particles are then made to collide according to the prescriptions of the
dynamics.  It is clear that in such a scheme the most time-consuming operation
is the search of the minimal collision time.  In one dimension, the number of
possible collisions between $N$~particles is~$N-1$, because the set of
positions is well-ordered.  This seems to imply immediately an ${\cal
O}(N)$~operations count for each collision.  Indeed, if we order the set of
collision times, finding the minimum takes ${\cal O}(1)$~operations, but
inserting a new collision time in the list will take ${\cal O}(N)$~operations. 
On the other hand, if we keep the set unordered, adding a new element will take
${\cal O}(1)$~operations, but finding the minimum will take~${\cal O}(N)$.  The
essence of an efficient algorithm is to use a data structure that
simultaneously permits fast insertion and fast search of the minimum.  This is
exactly the aim of the heap structure, well known in algorithmic
design~\cite{Kingston,knuth,sedge}.  Although known since a rather long time,
it is only recently that the heap concept has been used in physical problems,
like front propagation~\cite{sethian} or molecular dynamics simulations of
hard-sphere systems~\cite{marin1,marin,rapaport}.  In this paper we extend this
technique to systems with force fields acting between collisions, provided
these fields are Lagrangian invariant, or quasi-invariant.

The paper is organized as follows.  In section~\ref{s:heap} we introduce the
concept of a heap~\cite{Kingston}, and discuss code speed implementation
issues.  Section~\ref{s:algo} shows how an efficient event-driven evolution
scheme can be implemented by using the heap structure.  In
section~\ref{s:models}, we apply this code to the numerical solution of two
different one-dimensional systems of particles\,: the free streaming motion,
i.e. the evolution of $N$~non-interacting particles, and a classical
self-gravitating system~\cite{Miller,sev,tsu}.  This section also contains
checks on the speed of the algorithm.  In the conclusion section, we sum up our
results and discuss future applications.

\section{The heap}
\label{s:heap}

\noindent The discussion in this section mainly follows~\cite{Kingston}, the
classical references on the subject being~\cite{knuth,sedge}.

The {\em heap\/} structure is based on the concept of a tree.  A tree is
formally defined to be a set of {\em nodes}, connected by {\em links}, such
that the path to go from one node to another is unique.  A {\em rooted tree\/}
is a tree with distinguished node, {\em the root}, which is usually by
convention drawn on the top.  A rooted tree can be defined recursively as
follows\,: a rooted tree consists of a root node, and a finite set of subtrees,
which are themselves rooted trees.  Every node has a unique {\em parent}, except
the root.  The {\em children\/} of a node~$i$ are the roots of the subtrees,
under~$i$.  A node which has no children is called a {\em leaf}, and is usually
drawn at the bottom.  Binary trees are the most commonly used\,: they consist
of a root node and at most two subtrees which are themselves binary trees.

\begin{figure}[tb]
 \centerline{\psfig{file=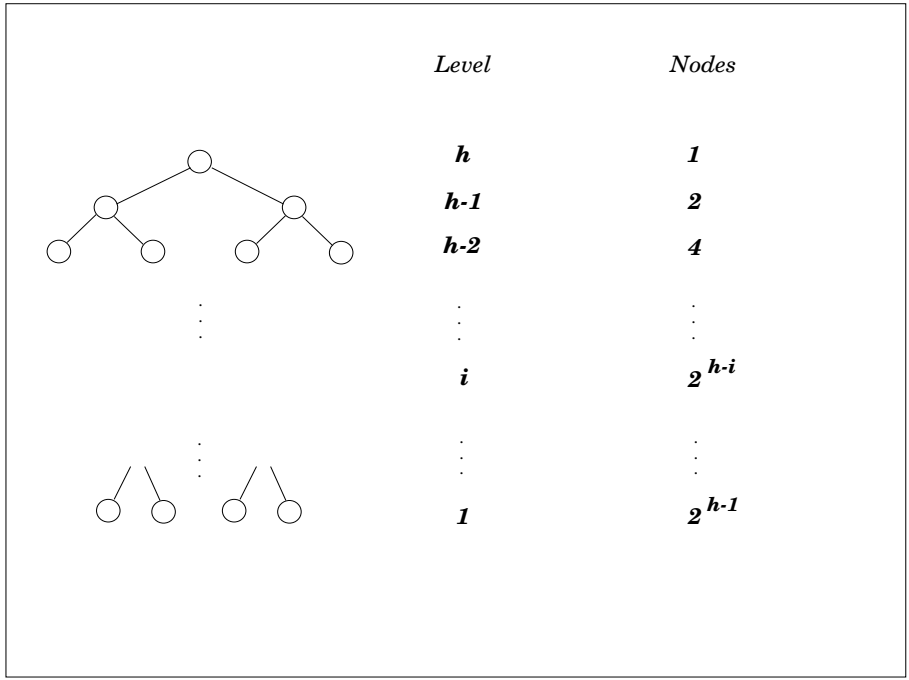,height=9cm,width=12cm}}
 \vspace{4mm}
 \caption{Example of an ordered complete binary tree.}
 \label{heapfig}
\end{figure}

A binary tree of $N$~elements is labeled ~$T(N)$.  The nodes can be graded in
levels.  The children of a node on level~$l$ are on level~$l-1$.  The
height~$h$ is a lower bound on the number of levels needed to build a binary
tree of $N$~elements, with~$h=\lceil\log_2 N\rceil+1$, as is illustrated in
figure~\ref{heapfig}.  In the following we will use {\em complete\/} and {\em
left-justified\/} binary trees, which are such that all the leaves are on
level~$2$ or~$1$, and those on level~$1$ are drawn from the left.  The nodes in
a complete and left-justified binary tree can be numbered from the top to the
bottom, and left to right, such that the children of node~$i$ are
respectively~$2i$ and~$2i+1$, while its parent (if any) is~$\lfloor
i/2\rfloor$.  Such a tree can be represented by an array with the array index
equal to the node number in the tree.

In a {\em heap-ordered\/} tree, an element is associated to each node, and the
elements obey the {\em heap condition}\,: each element in a child node is
greater than or equal to the elements in its parent node.  An array
representing a tree satisfying the heap condition is called a {\em
heap}~\cite{knuth,sedge}.

Given $N$~objects, the following strategy can be used to put them in a heap\,:
the first element is placed on the top of the tree, which corresponds to the
first element of the array.  The next element is placed in the second position,
and after comparison eventually replaces the first, if smaller.  This operation
is called {\em sift-up\/}~\cite{sedge}
 \footnote{The inverse operation ({\em sift-down\/}) makes an element move down
through the tree (percolate) until the heap condition is restored.  This
operation is typically needed when the heap has to be re-ordered after
replacement of an element.}.

This procedure is iterated\,: a new entry is placed at the end of the array,
and a comparison with its parent is performed.  If smaller, the element is
exchanged with the parent and is again compared to its new parent, and so on. 
As a result, each new element moves on an ascending path on the tree until the
heap condition is satisfied.

The above described simple procedure to initially order an array in a heap
(bottom-up) takes ${\cal O}(N\log N)$~operations at the most.  A faster scheme
(top-down), that only takes ${\cal O}(N)$~operations in the worst case is
however possible~\cite{knuth,sedge}.  Moreover, once the heap is built, the
operational cost of inserting or substituting a new element in a heap is at
most~${\cal O}(\log N)$, the height of the tree, and the selection of the
minimum is a trivial ${\cal O}(1)$~operation.  The heap is therefore a well
adapted data structure for both finding a minimum and replacing elements in an
array.

In many applications, elements have to be sorted according to different
criteria.  It is thus not possible to put them in a single heap, and it would
be very inefficient to duplicate data in different heaps.  The solution to this
problem is to use a single instance of all elements (that might already be
sorted itself according to one criterion), and to use {\em indirect\/} heap(s)
containing only pointers to them~\cite{sedge}.  When comparing elements while
moving through the heap, we access them through the (current) heap pointers
and, if the elements don't satisfy the heap condition, we simply exchange their
pointers without moving the elements themselves.  This procedure can be very
efficient if elements carry lots of related data, but it leads to a large
amount of memory traffic while moving through the heap, because of the need to
indirect through the pointers (but see below).

The concept of trees, and thus of heaps, can be generalized to bases larger
than two~\cite{lamarca}.  In a base~$r$ tree, each node has at most
$r$~subtrees so that, in a complete left-justified $r$-ary~tree, the children
of node~$i$ are $ri+2-r,\ldots,ri+1$ while the parent of node~$i$ is
$\lfloor(i-2+r)/r\rfloor = \lceil(i-1)/r\rceil$.  It is now clear that we face
two conflicting requirements\,: on one hand, we would like to minimize the
height of the tree~$h=\log_r N$ by choosing~$r$ as large as possible\,; on the
other hand, the work needed at each level of the tree to find the smallest of
the children, e.g. in a sift-down, increases linearly with~$r$, so we would
like to keep it small.  Blind minimization of the expression~$r\log_r N$
suggests that the best branching ratio would be~$e$, the base of natural
logarithms.  However, the processing of each level also incurs some work
independent of~$r$, and it is thus better to choose some higher (integer)
value.  This becomes even more important on modern microprocessors that access
memory through {\em caches\/} which are filled in bursts of typically 4, 8, or
16~words.  Because the children of a node are stored consecutively in memory,
it is then possible to use the fetching of a cache line to load all children of
one node at the same time.  This however also requires the heap to be {\em
cache-aligned\/}~\cite{lamarca}, which can be realized by fiddling with the
base address returned by the memory allocation routines in a language like~C. 
On the microprocessors we used (Alpha, Pentium, MIPS), we found that base~4 was
much better than base~2, while bases~8 and higher were slightly slower than
base~4.  The gain in speed by using aligned base~4 heaps with respect to
unaligned base~2 ones is significant, about~25\,\% on a Pentium with 15\,\%
coming from the choice of base and 10\,\% from the memory alignment.

To get all the benefits of aligned large base heaps, the comparison keys have
to be really present in the heap and not accessed through pointers that would
incur extra memory loads.  We thus implemented {\em semi-indirect\/} heaps in
which the keys are placed inside of the heap, so they can be compared directly,
while pointers to the corresponding elements are located in an array parallel
to the heap.  Because pointers have to be exchanged only when doing a swap,
this implementation reduces nearly by half the number of memory accesses (to at
most $r+1$~memory loads and 3~memory writes if a swap occurs in a sift-down)
and is faster than other priority queue implentations like those decribed
in~\cite{marin1}.

\section{The algorithm}
\label{s:algo}

\noindent We consider the motion of $N$~colliding particles in a
one-dimensional medium.  The interaction is not specified at this level\,: we
only require that the equation of motion for a particle can be integrated in
between two successive collisions.  Arrays of size~$N$ contain the states of
the particles, such as position, velocity and acceleration, at the time of
their last collision, stored in increasing order of the spatial coordinates. 
An additional state variable associated to each particle is~$\tau_j$, the time
it last experienced a collision.  Initially all~$\tau_j$ are set to zero.

\begin{figure}[tb]
 \centerline{\psfig{file=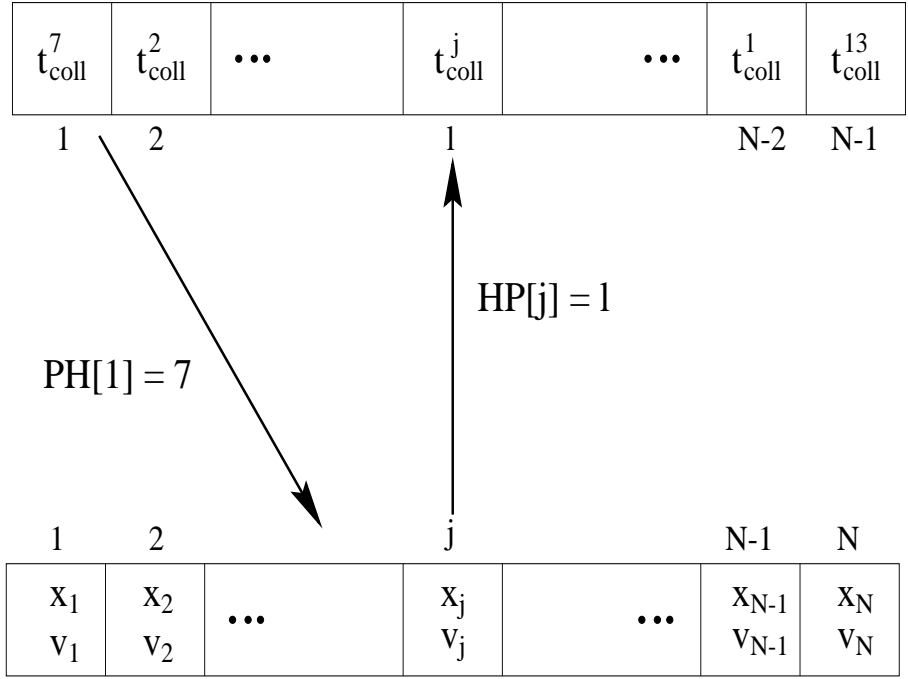,height=9cm,width=12cm}}
 \vspace{4mm}
 \caption{This figure shows the structure of a semi-indirect heap and the
function of the two ``shuffling'' arrays~$PH[\bullet]$ and~$HP[\bullet]$.  The
first array in the figure only contains the predicted collision times ordered
as a heap, while the second contains the particle states stored in increasing
order of spatial positions.  The two indexing arrays allow to move back and
forth between the two sets.}
 \label{arrayfig}
\end{figure}

The algorithm starts by computing the collision time of each particle with its
neighbor to the right, and the results are stored in an array of size~$N-1$. 
This array is then turned into a heap,~$T(N-1)$.  So that we do not need to
move the whole states of particles while processing the heap, we introduce an
indexing array, Particle-Heap~($PH[\bullet]$), mapping the position in the heap
to the position in space.  Referring to figure~\ref{arrayfig} if index~$l$
labels the position of the collision time in the heap, then $j=PH[l]$~is the
index in space of the leftmost of the two particles ($j$~and~$j+1$) involved in
that collision.  To update the list of predicted collision times of neighbors
particles, we also need the index array inverse to Particle-Heap, which we call
Heap-Particle~($HP[\bullet]$).  Hence for all~$j$ in the range 1~to~$N-1$
\begin{equation}
 PH[HP[j]] = j \quad\qquad\hbox{and}\quad\qquad HP[PH[j]] = j\ .
\end{equation}
This condition will be preserved at all times while we update the heap.  Note
that the collison times are really directly present in the heap, and that the
two indexing arrays then realize exactly the functions needed to implement the
semi-indirect heap.

The initial forming of the heap requires ${\cal O}(N)$~operations, as stated in
the previous section.  Once the heap has been built, the minimum collision
time~$t_{\rm min}$ is at the root.  The particles involved in the first
collision, which are $j=PH[1]$~and~$j+1$, are selected, and their states
evolved up to~$t_{\rm min}$.  The two particles are then at the same spatial
position and their states are rearranged by the collision (momenta simply
exchanged in the case of elastic collision).  The times~$\tau_j$
and~$\tau_{j+1}$, are set equal to~$t_{\rm min}$.  This procedure conserves the
monotonicity of the particle positions both at and after the collision.  Next
the new predicted collision times between $j$~and~$j+1$ is computed and
replaces the one at the root of the tree.  The root might now not fulfill the
heap condition, so the array probably has to be re-arranged with a sift-down of
the root value.  We need to check whether the new value is still less than the
values in its children.  If this is not the case, it is percolated down the
tree until the heap condition is satisfied.  This procedure involves at most
${\cal O}(\log N)$ operations, as discussed previously.

\begin{figure}[tb]
 \centerline{\psfig{file=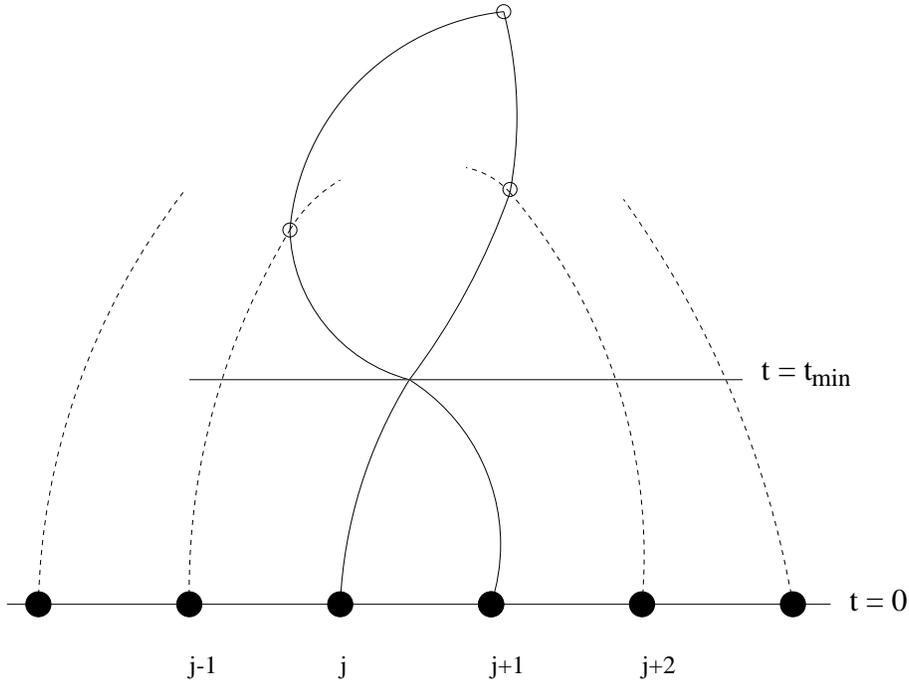,height=9cm,width=12cm}}
 \vspace{4mm}
 \caption{Intersection of the trajectories of particles~$j$ and~$j+1$ at
time~$t=t_{\rm min}$.  The ringed intersections are the collision/crossings
that need to be recomputed.}
 \label{collfig}
\end{figure}

Because of the changes of the states of particles $j$~and~$j+1$, their
collisions with their other nearest neighbors, $j-1$~and~$j+2$ need to be
re-computed, see figure~\ref{collfig}.  To do this, particles~$j-1$ and~$j+2$
are temporarily moved forward in time up to~$t_{\rm min}$, where their new
collision times with, respectively, particles~$j$ and~$j+1$ are computed, and
put into the heap, replacing the old ones.  As a consequence, the heap has to
be re-arranged two more times, again at a cost of at most~${\cal O}(\log N)$
for each modification.

The heap is now again in a consistent state, with the next collision time at
the root, and the whole procedure can be repeated.  The evolution can be
stopped either after some fixed number of collisions~$Z$, or when the predicted
time for the next collision becomes larger than some chosen final time~$T_{\rm
end}$.  In the end, all particles are moved forward in time from their
own~$\tau_j$ to the final time which is either~$T_{\rm end}$ or the time of the
last collision.  In conclusion, the complexity of the algorithm is in the
worst-case~${\cal O}(Z\log N)$ plus lower-order terms~${\cal O}(Z)$ and~${\cal
O}(N)$.

\section{Applications}
\label{s:models}

\noindent In this section we discuss two applications which we have used for
testing the performance of the algorithm.  In the first, we consider $N$~freely
moving particles in one spatial dimension.  The second problem we investigate
is the evolution of a Newtonian self-gravitating system
in~1-$D$~\cite{Dawson,Feix,Miller,tsu}.

\subsection{Free motion}
\label{free}

\begin{figure}[tb]
 \centerline{\psfig{file=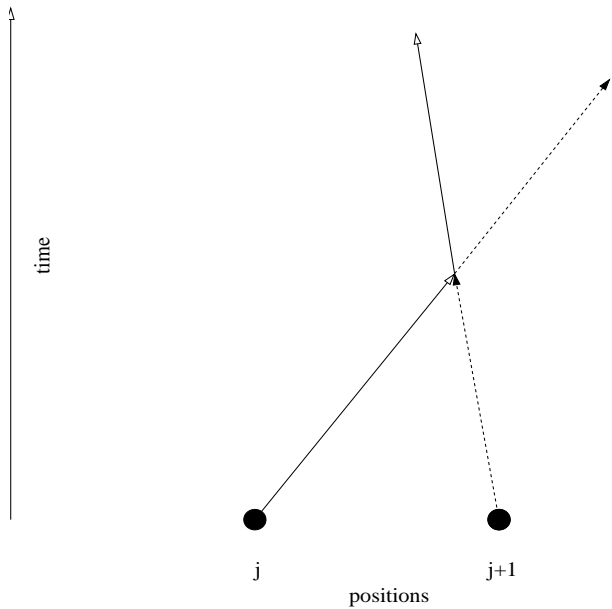,height=8cm,width=8cm}}
 \vspace{4mm}
 \caption{An elastic collision in the $(x,t)$~plane\,: the solid line traces
the motion of particle~$j$, while the dashed line shows particle~$j+1$.}
 \label{freefig}
\end{figure}

\noindent Consider $N$~particles all of the same mass~$m$ normalized to
be~$N^{-1}$.  Denote the position and the velocity of the $j$th~particle as
respectively $x_j$~and~$v_j$.  The particles move freely, hence neither
interact with each other, nor do they feel any external force.  Their
acceleration is thus identically equal to zero.  In the plane~$(x,t)$, each
particle moves on a straight line, the slope of which is its velocity.  The
intersection of two lines represents a crossing or a collision in physical
space.  Each time two particles encounter, i.e.~$x_j=x_{j+1}$, they cross each
other.  Equivalently, collisions can be thought of as elastic, since
one-dimensional motion of indistinguishable particles is equivalent to a system
of one-dimensional impenetrable mass points, which bounce elastically off one
another (see Fig.~\ref{freefig}).  In the latter case, colliding particles
exchange their velocities\,: in the $(x,t)$~plane the trajectory of a single
particle is then a broken line.

We remark that the system of non-interacting particles is not trivial in
Eulerian coordinates, and has been used as a model of structure formation in
the early Universe, the so-called Zeldovich approximation.  Furthermore, as
long as the solution stays single-stream, i.e. before any collision has
occurred, the Zeldovich approximation is in some sense exact in one
dimension~\cite{massimo,gurbatov}.  After a first collision, when in the
continuum limit a caustic has been formed and the solution is no longer
single-stream, the dynamics of a self-gravitating system can nevertheless be
proven to stay close to the Zeldovich approximation for short enough
times~\cite{royt}.

\begin{figure}[tb]
 \centerline{\psfig{file=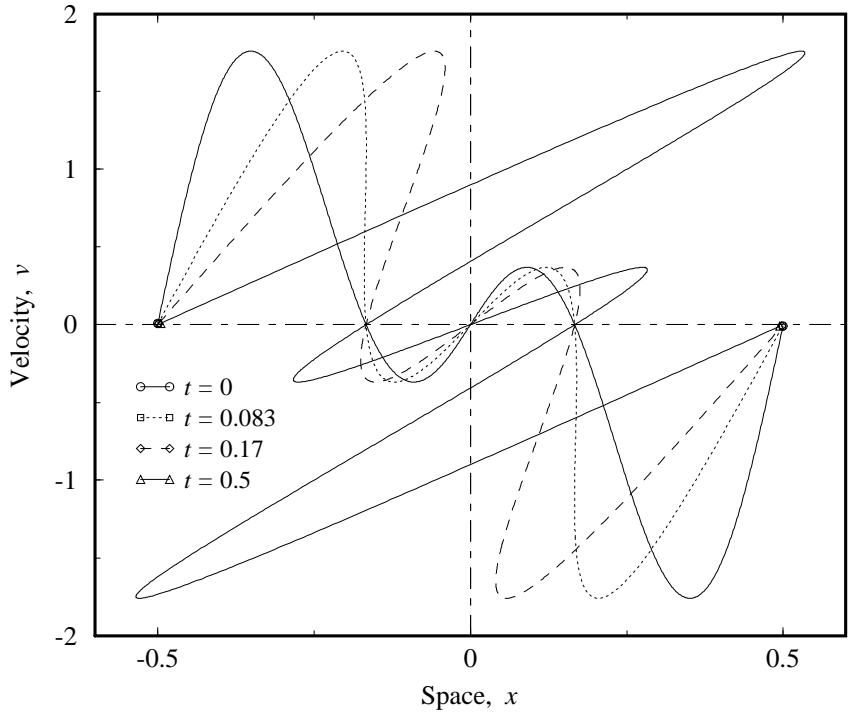,width=10cm}}
 \vspace{12mm}
 \caption{Phase space portraits of the free motion starting from initial
velocity on double sine wave.  The first caustic is formed at
time~$t=(4\pi)^{-1}$ (dotted line).  After caustic formation, velocity is a
multi-valued function of position and in that region, a Zeldovich pancake, or
blini, carries an increasing fraction of the mass.}
 \label{space1}
\end{figure}
\begin{figure}[tb]
 \centerline{\psfig{file=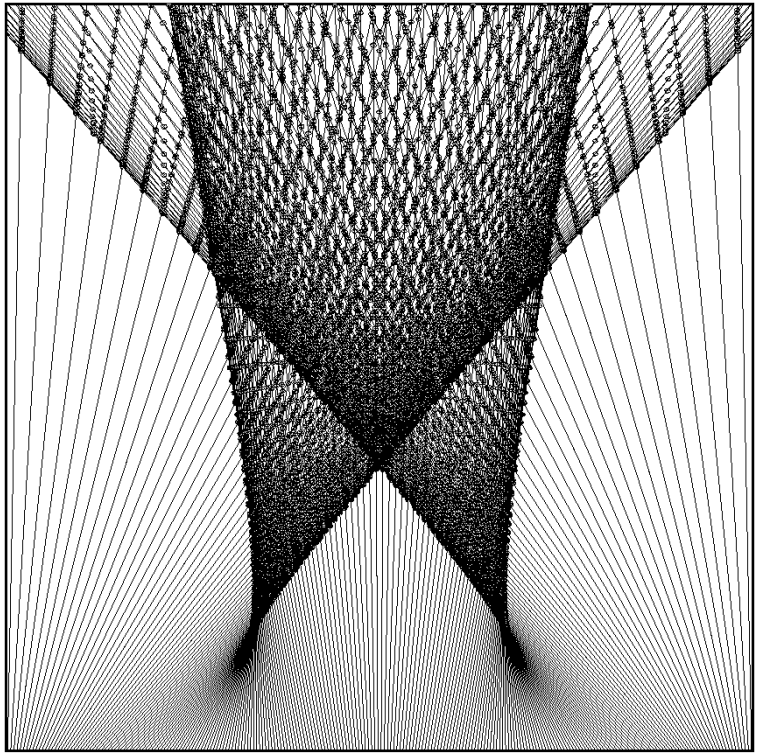,width=10cm,height=10cm}}
 \vspace{4mm}
 \caption{Dynamics in the $x,t$~plane for free motion with the same initial
conditions as Fig.~\ref{space1}.  The two caustics are clearly visible.  For
long times, all particles will fly away to infinity.}
 \label{xt1}
\end{figure}

Fig.~\ref{space1} shows a phase-space portrait of this dynamics with particles
initially uniformly distributed in space, and velocity a smooth function of
position.  Fig.~\ref{xt1} shows the same dynamics in the $(x,t)$~plane, and
clearly displays structure formation in Eulerian coordinates (caustics).

\subsection{Self-gravitating systems}
\label{grav}

\noindent Consider now a one-dimensional (classical) Newtonian self-gravitating
system of $N$~particles, again all of the same mass~$m=N^{-1}$.  The
Hamiltonian is\,:
\begin{equation}
 H = \sum_{j=1}^N {p_j^2 \over 2m}
+2\pi G m^2 \sum_{j=1}^N\sum_{i>j}^N |x_i - x_j|\ ,
 \label{Ham}
\end{equation}
where $x_j$~is the position of particle~$j$, $p_j\equiv m v_j$~is the momentum
conjugate to~$x_j$ and $G$~is the gravitational constant~\cite{Miller,tsu}.  We
choose as unit of length the spatial interval in which the particles are
initially contained.  The initial density~$\rho_0$ is thus equal to one.  The
natural choice of time scale is the inverse of the Jeans frequency~$\omega_{\rm
J} = (4\pi G\rho_0)^{1/2}$.  With our choice of length this implies that we
take~$4\pi G$ equal to~one.

Inbetween two collisions, the acceleration of each particle is constant, and is
proportional to the difference of number of particles on its right and on its
left.  In the $(x,t)$~plane, the path of a particle between collisions thus
follows a parabola.  If the particles are assumed to pass through each other
freely, then in a collision, velocities are unchanged but accelerations are
exchanged.  If, on the other hand, particles are assumed to scatter
elastically, then in a collision, velocities are exchanged but accelerations
are unchanged.  Switching from one interpretation to the other only involves
keeping track of the permutation relating the current rank of the particles to
their initial rank.  This can be realized algorithmically by bookkeeping at
each collision two indexing arrays, inverse of each other in a way similar to
the~$HP[\bullet]$ and~$PH[\bullet]$ arrays, but this time holding the relations
between the initial and the current particle rank.

Evolving the system thus involves basically two operations\,: finding the next
collision time of a pair of particles and moving these to their common
collision time.  Although seemingly simple, these two operations contain lots
of numerical traps.  We remark that the system is chaotic, i.e. dynamically
unstable, and amplifies small perturbations.  It is thus especially important
to keep numerical errors small.  First, finding the collision time implies
finding the positive root of the quadratic equation
\begin{equation}
 {1 \over 2}\delta a(t_{\rm c}-t)^2+\delta v(t_{\rm c}-t)+\delta x = 0
 \label{equadr}
\end{equation}
where~$\delta a$, $\delta v$~and~$\delta x$ are respectively the differences of
accelerations, velocities and positions between the right and left particles at
the common time~$t$ (so that~$\delta x > 0$ and~$\delta a < 0$), and $t_{\rm
c}$~is their predicted next collision time.  That both roots are real and that
there can be only one positive root follows from the fact that~$\delta a\delta
x$ is negative.  As is well known, solving the quadratic equation by the
classical formula\,:
\begin{equation}
 t_{\rm c}-t = {-\delta v\pm\sqrt{\delta v^2-2\delta a\delta x} \over \delta a}
 \label{sol1}
\end{equation}
will lead to severe cancellation when $\delta v$~is negative, $\delta x$~is
small and the positive root is sought, that is exactly the case that
corresponds physically to nearly free motion.  Hence we only use~(\ref{sol1})
when $\delta v$~is positive and with the $-$~determination of the radical. 
When $\delta v$~is negative, the classical formula can be re-arranged to\,:
\begin{equation}
 t_{\rm c}-t = {2\delta x \over -\delta v+\sqrt{\delta v^2-2\delta a\delta x}}
 \label{sol2}
\end{equation}
which is stable in this case.  Note that both formulas naturally tend to their
proper limit when~$\delta a \rightarrow 0$ (the free motion case).  In the
self-gravitating case and with our choice of units, $\delta a$~has the constant
value~$-N^{-1}$ for neighboring particles.

Second, moving the particles forward in time should be written as
\begin{equation}
 x_j(t) = x_j+\left[v_j+{a_j \over 2}\left(t-\tau_j\right)\right]
\left(t-\tau_j\right)
 \label{sol3}
\end{equation}
that, if~$a_j/2$ is precomputed, involves two nested {\em fused
multiply-add\/}~({\tt madd}) operations, which on many modern microprocessor
are executed in a single clock cycle and with a single roundoff error, leading
to a saving of three floating-point operations over the five needed by the
classical expression $x_j+v_j(t-\tau_j)+a_j/2(t-\tau_j)^2$. 
Expression~(\ref{sol3}) can also be recognized as {\em Horner's rule}, which,
even on machines lacking a {\tt madd}~instruction, saves one multiplication.

If the positions of two colliding particles are updated according to
equation~(\ref{sol3}), they might end up with a slightly different final
position or, even worse, the left particle could overtake the right one because
of roundoff.  To prevent this we used a common symmetric formula for computing
the final position of the pair,
 \begin{eqnarray}\kern -2em
 x_{\rm c}(t_{\rm c})&=&\phantom{{}+{}}{x_j+x_{j+1} \over 2} \nns
 &&{}+\left[v_j-v_{j+1}
 +\left({a_j+a_{j+1} \over 2}\right)\left({\tau_{j+1}-\tau_j \over 2}\right)
 \right. \nns &&\left.\phantom{{}+\left[v_j-v_{j+1}\right.}
 +(a_j-a_{j+1})\left(t_{\rm c}-{\tau_{j+1}+\tau_j \over 2}\right)\right]
 \left({\tau_{j+1}-\tau_j \over 2}\right){1 \over 2} \nns
 &&{}+\left[v_j+v_{j+1}
 +\left({a_j+a_{j+1} \over 2}\right)
 \left(t_{\rm c}-{\tau_{j+1}+\tau_j \over 2}\right)\right]
 \left(t_{\rm c}-{\tau_{j+1}+\tau_j \over 2}\right){1 \over 2}
 \end{eqnarray}
which is physically equivalent to moving the center of mass of the two
particles.  Although seemingly complicated, this form is full of repeated
subexpressions and can thus be evaluated quite efficiently, especially as it
again only involves {\tt madd}~operations.  This expression also gives the
benefit of preserving exactly any symmetries that might be present in the
initial velocity profile because positions, velocities and accelerations are
combined {\em before\/} any multiplication is done.

\begin{figure}[tb]
 \centerline{\psfig{file=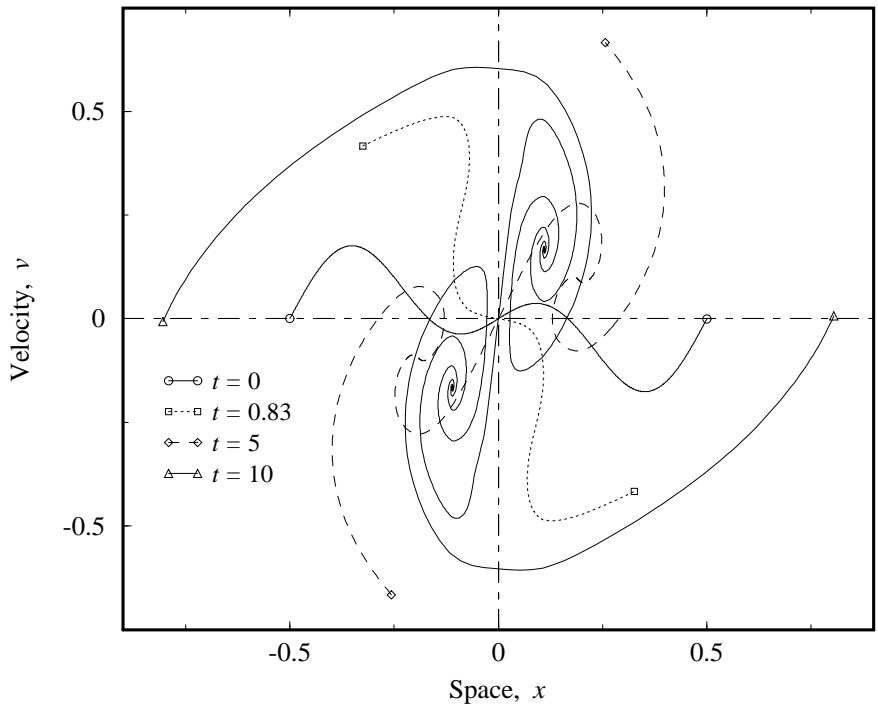,width=10cm}}
 \vspace{12mm}
 \caption{Phase space portrait of a self-gravitating system.  The initial
velocity profile is the same as in Fig.~\ref{space2}, i.e. a double sine wave. 
Although the initial evolution is similar to the free motion case, after a few
Jeans~times, the system develops spiral structures in phase space, due to the
gravitational attraction preventing the particles from running away.}
 \label{space2}
\end{figure}
\begin{figure}[tb]
 \centerline{\psfig{file=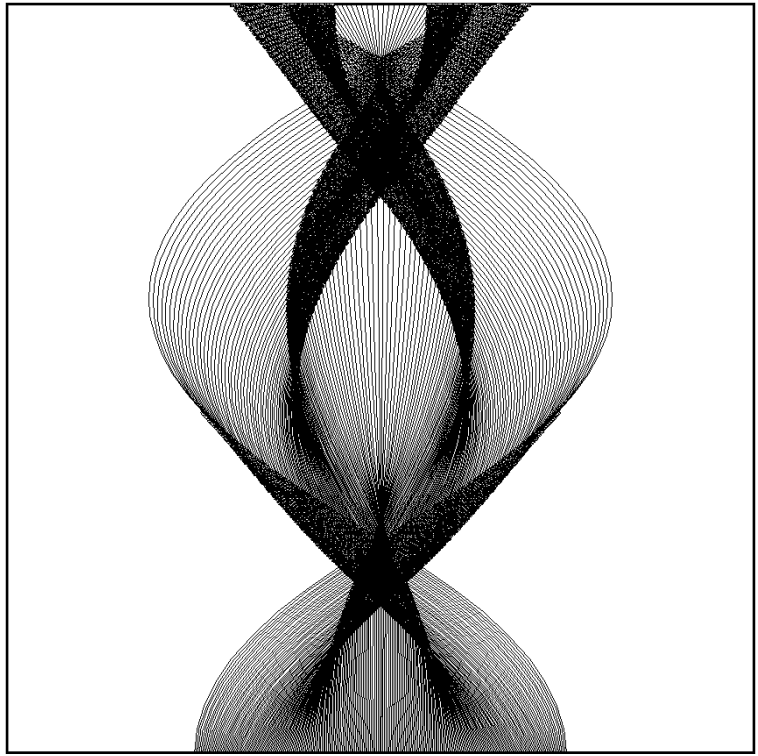,width=10cm,height=10cm}}
 \vspace{4mm}
 \caption{Dynamics in the $x,t$~plane for a self-gravitating system.  For large
times, all particles stay trapped in a finite region of space, with their
trajectories intermixing each other until the system becomes completely
chaotic.}
 \label{xt2}
\end{figure}

Fig.~\ref{space2} shows phase space portraits of this self-gravitating dynamics
with particles initially uniformly distributed in space, and velocity a smooth
function of position.  After caustic formation, the system develops a spiral
structure in phase-space.  Fig.~\ref{xt2} shows the same dynamics in the
$(x,t)$~plane.  As in free streaming motion, caustics can be observed.  In
addition, we see that gravitation stops the particles from moving apart as
easily after the caustics have been formed and this leads to a more
concentrated mass agglomeration.  Nevertheless, it is worth stressing the
qualitative agreement between Figs.~\ref{xt1} and~\ref{xt2} for the initial
stage of the evolution\,: this is an indirect confirmation of the validity of
the Zeldovich approximation.

An event-driven scheme for the simulation of one-dimensional self gravitating
systems was first introduced by Eldridge and Feix~\cite{Feix}, and has later
been further developed in the literature~\cite{sev}.  This and all other
published schemes however always involve at least ${\cal O}(N)$~operations per
collisions, either to find the minimum collision time, or to update the
particle states after each collision.  In our algorithm, these two operations
need respectively~${\cal O}(\log N)$ and~${\cal O}(1)$ operations.

\begin{figure}[tb]
 \centerline{\psfig{file=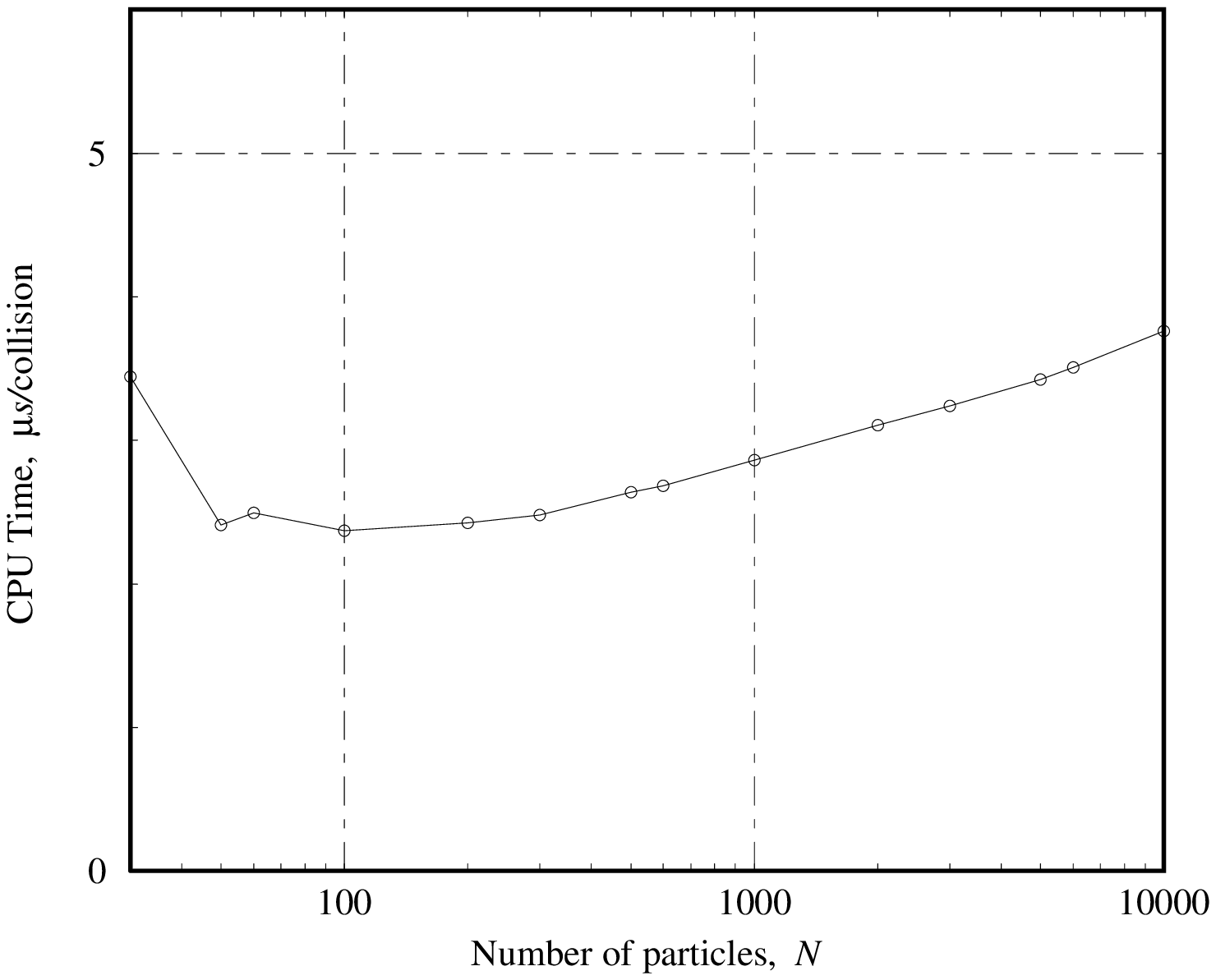,width=10cm}}
 \vspace{12mm}
 \caption{CPU~time per collision measured as the output of the UNIX library
function~{\tt times()}, divided by the number of collisions, after 30~Jeans
time, for systems of various size~$N$.  The code was compiled by~{\tt gcc} in
the Linux kernel~2.2 and ran on an Intel Pentium~II 450\,MHz processor.  Data
points on the left are not very reliable because of the limited resolution
of~{\tt times()}.}
 \label{cpu}
\end{figure}

The theoretical predictions of the performance of the algorithm is confirmed by
Fig.~\ref{cpu}, which shows CPU time per collision {\it vs.\/}~number of
particles in semi-logarithmic scale.  The linear dependence on~$\log N$ is
clear in the range~300--10000.  However, there is a significant constant
contribution coming from the floating point operations needed to update the
particles states, which actually still take the lion's share of the CPU time
for 10000~particles.  In the data of Fig.~\ref{cpu} (see caption), we reached
speeds in excess of $4\,\times\,10^5$\,collisions/s on an inexpensive Linux~PC.
On a DEC ALPHA-based workstation at 600\,MHz, we got speeds of around
$1.3\,\times\,10^6$\,collisions/s, for $N$~equal to~1000.  Hence the algorithm
presented here allows the study of fairly large systems for long times on
ubiquitous hardware.

\begin{figure}[tb]
 \centerline{\psfig{file=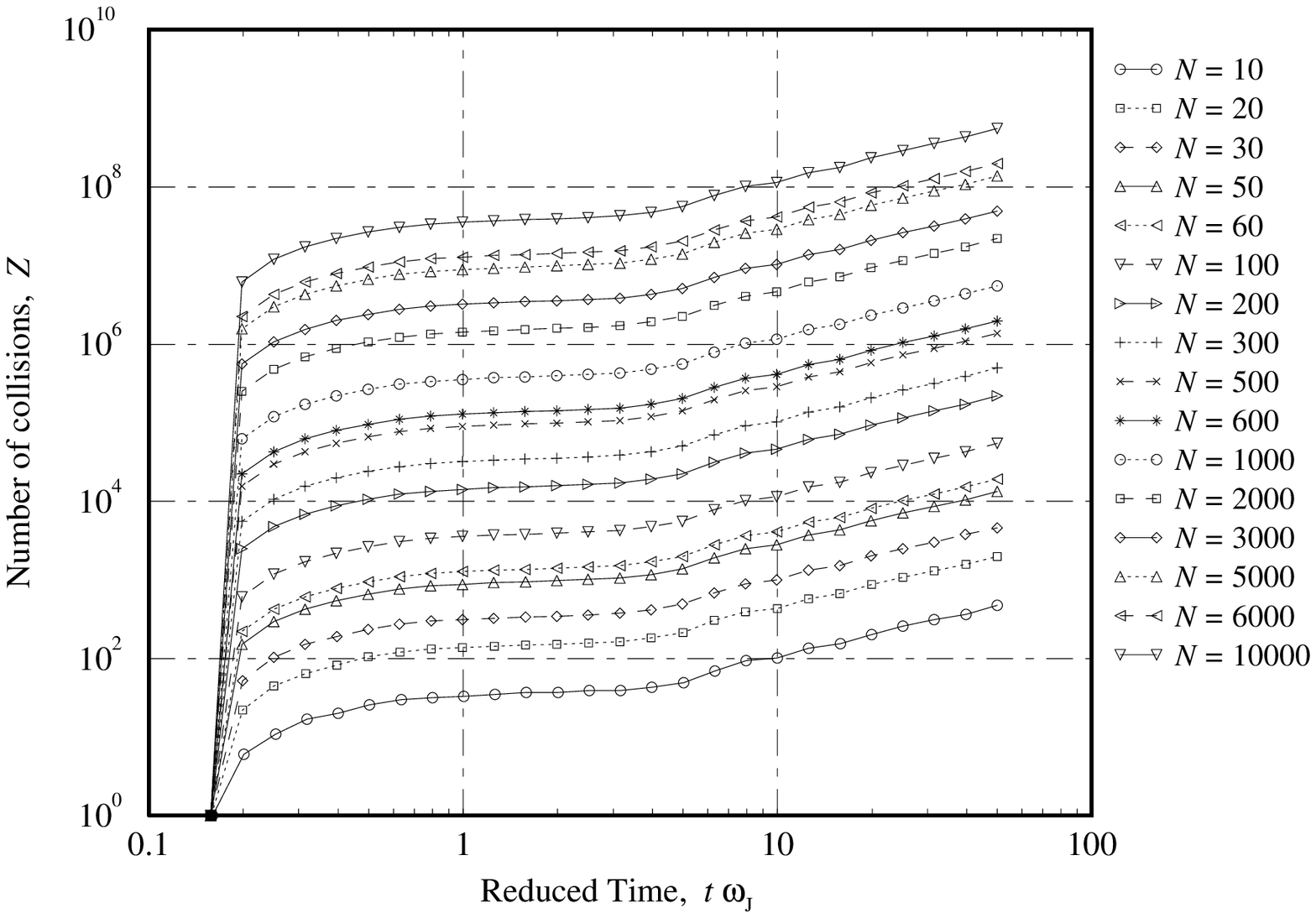,width=10cm}}
 \vspace{12mm}
 \caption{Number of collisions {\em vs.\/}~time in units of~$\omega_{\rm
J}^{-1}$ for different number of particles.  We can clearly distinguish one
early regime, before many collisions have occurred, one late regime (after
about~$10 \omega_{\rm J}^{-1}$) where the collision rate becomes constant, and
one intermediate regime.  For all times, the number of collisions goes
as~$N^2$.}
 \label{log1}
\end{figure}
\begin{figure}[tb]
 \centerline{\psfig{file=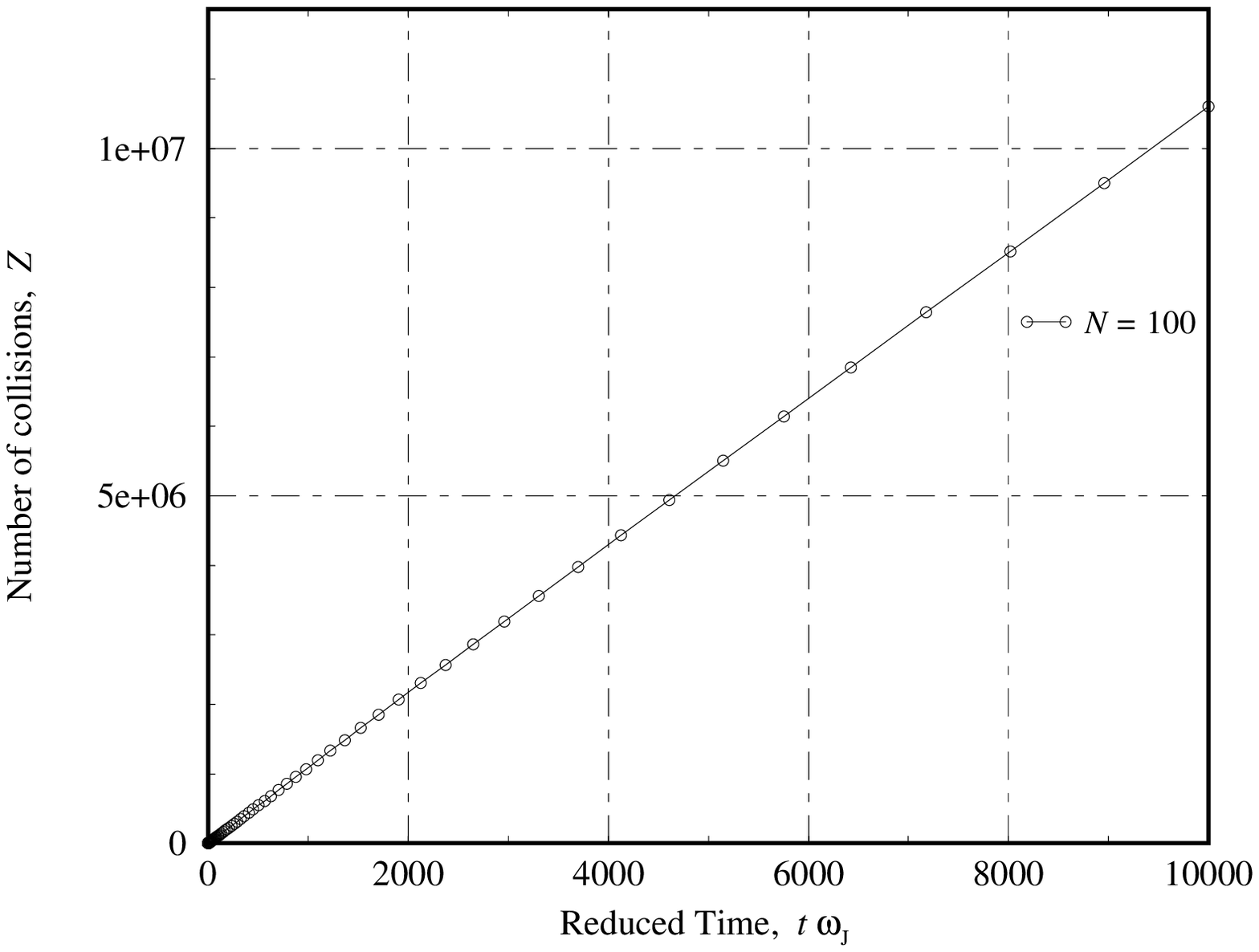,width=10cm}}
 \vspace{12mm}
 \caption{A different representation of Fig.~\ref{log1}, for only one value
of~$N$, but for much longer times and in linear coordinates.  The collision
rate has clearly become constant at all times displayed in this figure.}
 \label{log2}
\end{figure}

The preceding discussion has made clear that a limiting factor for simulating a
system up to a given time~$T_{\rm end}$ is the number of collisions~$Z$ that
have to be performed in that time interval.  If we want to approach the
continuum limit, that is large~$N$\,s, it is especially important to know how
$Z$~scales with~$N$.  If $N$~mass points with total mass equal to one,
initially located in an interval of length~one, provide a discretization of a
given velocity function, then, as long as the discretization is not felt, one
expects that the average time between successive collisions of a given particle
goes like~$N^{-1}$.  The total collision rate hence grows as~$N^2$.  Likewise,
if we consider discretization of a statistically stationary state, the distance
between particles would scale as~$N^{-1}$ while the velocity would be
independent of~$N$.  Hence, also in this case, the collision rate would be
proportional to~$N^2$.  Fig.~\ref{log1} shows the number of collisions {\it
vs.\/}~time in unit of inverse Jeans' frequency, in double logarithmic
coordinates, for different number of particles.  The curves corresponding to
different~$N$\,s are parallel to each other, the separations being the squares
of the ratio of successive values of~$N$.  Fig.~\ref{log1} hence suggests that
the proposed scaling holds true for different discretizations of the same
initial conditions for all regimes.  Fig.~\ref{log2} explores the late time
regime, and shows the number of collisions {\it vs.\/}~time in linear scale
(the earlier time regimes shown in Fig.~\ref{log1} are not visible in this
representation).  It is clear from this figure that the collision rate becomes
constant for times much larger than the inverse Jeans' time.

\section{Conclusions}

\noindent We discussed the implementation of a fast heap-based event-driven
scheme for integrating numerically one dimensional systems of $N$~interacting
particles, provided the dynamics can be integrated between two successive
collisions.  The collision times are ordered on a heap, which reduces the
complexity to ${\cal O}(\log N)$~operations per collision.  As a consequence,
for large values of~$N$, the present algorithm is faster than earlier
algorithms in the litterature, which are~${\cal O}(N)$.  This opens up the
perspective of improving the numerical estimates of the statistics of such
systems.

Commonly, a particle system is considered as a discrete approximation to a
continuum limit, e.g. self-gravitating particles to the Vlasov-Poisson system
of coupled partial differential equations.  If so, the main limitation of the
present scheme is~$Z$, the number of collisions needed to be performed in a
given (intrinsic) time~$T_{\rm end}$.  We presented theoretical arguments and
numerical simulations showing that $Z$~generally grows as~$N^2$.  Hence the
total computational cost of reaching a time~$T_{\rm end}$ is~${\cal O}(g(T_{\rm
end})N^2\log N)$, where $g$~is a function depending also on the initial
condition, but which becomes proportional to~$T_{\rm end}$ for sufficiently
large~$T_{\rm end}$, i.e.~$g\approx f\,T_{\rm end}$ with $f$~independent of~$N$
and~${\cal O}(1)$ if~$v$ is~${\cal O}(1)$ (see Fig.~\ref{log2}).  We believe
that one cannot do better for one-dimensional particle systems than the
algorithm presented here, without introducing further approximations.

For many applications, such as self-gravitating particles, the system is
inherently chaotic, and small errors are amplified by the dynamics.  Given some
initial accuracy~$\varepsilon$, the best accuracy that can be got at
time~$T_{\rm end}$ is~$\varepsilon\,e^{\lambda T_{\rm end}}$, for some
positive~$\lambda$.  Given this, we might however estimate the relative errors
of discretization of a given~PDE to a particle system, and the roundoff errors
from the collisions, in order to find the ``best'' value of~$N$ leading to the
smallest total error.  The first error is essentially~$N^{-1}$, while the
second is~$\eta(f\,T_{\rm end} Z/N)^{1/2}$, where $\eta$~is the machine
precision and $f\,T_{\rm end} Z/N$~is the average number of collisions
experienced by a single particle in the interval~$[0,T_{\rm end}]$, and the
roundoff errors coming from each collision being uncorrelated.  The smallest
total error is hence found for $N$~of the order of~$\eta^{-2/3}(f\,T_{\rm
end})^{-1/3}$ and would thus scale as~$\eta^{2/3}(f\,T_{\rm
end})^{1/3}\,e^{\lambda T_{\rm end}}$.  We see from these expressions that, as
expected, the optimal value of~$N$ decreases with~$T_{\rm end}$, and that it is
useless to use more than~$\approx 10^4$ particles in single precision and more
than~$\approx 10^{10}$ in double precision.  While the second number of
particles is out of reach of current computers, the first one can be handled by
our code (about 25~CPU~seconds per Jeans time for~$N=10000$).  In double
precision, we should always use values of~$N$ as large as possible as the
limiting value is very large and decreases only very slowly with~$T_{\rm end}$.

It should be added that in both applications presented here, the particle view
is at least as fundamental as the PDE one.  The discretization gives
coarse-grained noise, compared to the PDE, but this is a real physical effect
e.g. in stars clusters~\cite{henon,meylan}.  Furthermore, for investigation of
the statistically steady state in such system, the relevant errors on global
quantities are not growing with time in this regime, assuming the validity of
the shadowing lemma of dynamical systems theory.

In the paper, we presented free motion and self-gravitating systems as possible
applications of our algorithm.  Nevertheless, it is worth to stress that the
algorithm is more general and, for example, can also be applied to models of
the motion of matter in an expanding Universe~\cite{Aurell,Rouet1990}.

\section*{Acknowledgments}
\label{s:acknowledgements}

We thank U.~Frisch, M.~H\'enon and P.~Muratore-Ginanneschi for discussions.  We
also thank D.~Zanette for pointing out reference~\cite{marin1}.

This work was supported by RFBR-INTAS~95-IN-RU-0723 (E.A. and D.F.), by the
Swedish Natural Science Research Council through grants M-AA/FU/MA~01778-333
(E.A.) and M-AA/FU/MA~01778-334 (D.F).

\end{article}

\end{document}